\documentclass[prd, showpacs, preprint, floatfix]{revtex4}

\usepackage{latexsym}
\usepackage{amssymb}
\usepackage{amsmath}\large\huge\normalsize
\usepackage{graphicx}
\usepackage{amsfonts}
\usepackage{array}
\usepackage{subfigure}
\usepackage{float}
\usepackage{tabularx}

\begin{document}

\newcommand{\be}{\begin{equation}}
\newcommand{\ee}{\end{equation}}
\newcommand{\bea}{\begin{eqnarray}}
\newcommand{\eea}{\end{eqnarray}}

\newcommand{\comment}[1]{}

\title{
Dynamical Surface Gravity in Spherically Symmetric\\ Black Hole Formation }

\author{Mathias Pielahn}
\email[Electronic address: ]{mathiaspielahn@gmail.com}
\affiliation{Department of Physics, University of Winnipeg, Winnipeg, Manitoba, Canada, R3B 2E9}

\author{Gabor Kunstatter}
\email[Electronic address: ]{g.kunstatter@uwinnipeg.ca}
\affiliation{Department of Physics and Winnipeg Institute of Theoretical Physics, University of Winnipeg, Winnipeg, Manitoba, Canada, R3B 2E9}

\author{Alex B. Nielsen}
\email[Electronic address: ]{alex.nielsen@aei.mpg.de}
\affiliation{Max-Planck-Institut f\"{u}r Gravitationsphysik, Albert-Einstein-Institut, D-14476, Golm, Germany}
\date{\today}

\begin{abstract}
 We study dynamical surface gravity in a general spherically symmetric setting using Painlev\'{e}-Gullstrand (PG) coordinates. Our analysis includes several definitions that have been proposed in the past as well as two new definitions adapted to PG coordinates.  Various properties are considered, including general covariance, value at extremality, locality and static limit. We illustrate with specific examples of ``dirty" black holes that even for spacetimes possessing a global timelike Killing vector, local definitions of surface gravity can differ substantially from ``non-local" ones that require an asymptotic normalization condition. Finally, we present numerical calculations of dynamical surface gravity for black hole formation via spherically symmetric scalar field collapse. Our results highlight the differences between the various definitions in a dynamical setting and provide further  insight into the distinction between local and non-local definitions of surface gravity.
\end{abstract}

\pacs{04.25.dc, 04.70.Dy}

\maketitle

\section{Introduction}
An event horizon is a global property of a black hole spacetime that seems to require, in classical general relativity at least, the existence of a singularity. It is hoped that the singularity inside a black hole will ultimately be resolved in a complete quantum theory of gravity, but it is not clear what effect singularity resolution would have on the existence of event horizons. For example, the global event horizon may be replaced by a compact trapping horizon \cite{hayward2,ziprick2}. Most work on black hole thermodynamics deals with Killing horizons. These are idealizations whose existence is belied in all but the most contrived circumstances by the presence of Hawking radiation.  While such idealizations are very useful, it is ultimately necessary to compute corrections to the thermodynamics that arise from the dynamics of the horizon. The thermodynamic properties of black holes therefore need to be formulated in a more general setting that allows in principle for the growth by accretion and shrinking by radiation of the horizon. One of the first attempts to systematically describe such a dynamical setting for black hole event horizons was given in the mid-nineties by Hayward \cite{hayward1} who introduced the notion of a ``trapping horizon." More recently Ashtekar and collaborators formulated a rigorous definition of a ``dynamical" horizon \cite{ashtekar1}. 

Surface gravity is a classical construct that plays an important role in black hole thermodynamics. It is therefore somewhat surprising that a generally accepted definition of surface gravity only exists for Killing horizons in static or stationary spacetimes \cite{wald,nielsen1}. The standard definition is based on the existence of a global time translational Killing vector field that becomes null on the event horizon. In a dynamical setting the Killing vector ceases to exist so the standard (Killing) definition of surface gravity is no longer valid. Various dynamical definitions have been proposed over the years to address this problem \cite{collins,nielsen1,fodor,nielsen2,hayward,visser,abreu} but to the best of our knowledge no consensus exists on which is correct. A systematic analysis of the relative merits of some of these definitions was given by Nielsen and Yoon \cite{nielsen1}. The present paper builds on this analysis by presenting an analytic and numerical study of various dynamical definitions, including two new ones that have not previously been considered. 

In the first part of the paper we examine the various definitions analytically and classify them according to several fairly basic criteria.  The criteria we consider are: 
\begin{enumerate}
\item Is the definition coordinate invariant? 
\item Does the proposed definition vanish for extremal horizons? 
\item Is the definition local in the sense of not depending on the matter distribution outside the horizon? 
\item Does the definition in general reduce to the standard textbook (i.e. Killing) definition for all asymptotically flat black hole spacetimes with a global Killing vector? 
\end{enumerate}
We will see that only two of the five definitions we consider answer 2. in the affirmative. Moreover, if the answer to 3. is yes then the answer to 4. must be no. 

The second part of the paper presents a numerical study of dynamical surface gravity for horizon formation via the collapse of a spherically symmetric massless scalar field. The results demonstrate in a precise physical context the differences between the various definitions in the dynamical region, i.e. while the apparent horizon is growing. In addition, the simulation provides  a graphic illustration of the qualitative and quantitative difference between local and non-local definitions after the horizon has settled down to its final value.

We work primarily in generalized Painlev\'{e}-Gullstrand (PG) coordinates, which are one representative of a class of coordinates that are regular across future horizons. They are distinguished by the fact that spatial slices are everywhere flat. The PG metric takes the form:
\be
ds^2 = -\sigma(t,r)^2 dt^2 +\left( dr+\sqrt{\frac{2GM(t,r)}{r}}\sigma(r,t) dt \right)^2 + r^2 d\Omega^2, \label{eq:PG}
\ee
where \(M(t,r)\) is the generalized Misner-Sharp mass function \cite{misner_sharp}.
It has a physical interpretation as the quasi-local energy contained within a sphere of areal radius $r$ at time $t$. In asymptotically Schwarzschild spacetimes the Misner-Sharp mass approaches the ADM mass at spatial infinity, whereas in radiating spacetimes it reproduces the Bondi mass at future null infinity.  The lapse function $\sigma(t,r)$ approaches a function independent of $r$ in the asymptotically Schwarzschild region and can be set to one there. This will play an important role in the interpretation of the numerical results in what follows.

The paper is organized as follows: in the next Section we present the definitions of dynamical surface gravity that we will be studying and   evaluate them in PG coordinates in a general spherically symmetric setting. Section III shows that in the context of so-called ``dirty'' black holes (those with non-zero stress energy outside of static horizons) one must in general choose between criteria 3. and 4. given above. 
The extremality condition for dynamical horizons is discussed in Section IV while Section V describes our numerical methods and presents the results. The last Section contains a summary and conclusions.

\section{Definitions of Dynamical Surface Gravity}
While some of the following definitions are contained in the papers by Nielsen \cite{nielsen3}, and Nielsen and Yoon \cite{nielsen1}, we also propose two new definitions that are motivated by our focus on PG coordinates.

\subsection{Static Spacetimes: The Killing Definition}
When the mass function and lapse function depend only on the radial coordinate, the spacetime is stationary and possesses a timelike Killing vector, $t^\alpha$. The standard definition of surface gravity $\kappa_{Killing}$ for a static black hole with a Killing horizon \cite{wald} is:
\be
t^{\alpha} \nabla_{\alpha} t^{\beta} = t^{\beta} \kappa_{Killing},
\ee
 where the Killing vector is normalized to have unit norm at spatial infinity. Any static, spherically symmetric metric can be written in Schwarzschild coordinates as follows:
\be
ds^2=-\sigma(r)^2f(r)dt_{s}^2+f(r)^{-1}dr^2+r^2d\Omega^2,
\label{eq:schwarz static}
\ee
where $t_s$ is the Schwarzschild time. In the static case it is straightforward to transform to PG time $t$ in which the metric takes the form Eq.~(\ref{eq:PG})
with \(GM/r=1-f\) and both $M(r)$ and $\sigma(r)$ functions of $r$ only. Note that the transformation from Schwarzschild to PG time leaves the lapse unchanged. As well, we assume asymptotic flatness, so that both $M(r)\to m$ and $\sigma(r)\to \sigma_\infty$ go to constants at spatial infinity. With the suitably normalized global Killing vector \(t^{\alpha}=(1/\sigma_\infty,\,0,\,0,\,0)\) we have
\be
\kappa_{Killing} = \left(\frac{\sigma_{hor}}{\sigma_\infty}\right)\frac{1}{4GM_{hor}} \left(1-2GM^\prime \right)|_{hor},
\label{eq:comparison}
\ee
where the subscript \(hor\) means evaluation on the outer apparent horizon and \(^\prime\) denotes differentiation with respect to the radial coordinate. If one subscribes to the standard definition of surface gravity for static black holes, then all proposed definitions should coincide with (\ref{eq:comparison}) in the limit that the spacetime is static and the horizon is Killing. Without loss of generality $\sigma_\infty$ can be chosen to be unity and we will henceforth assume that this is the case unless otherwise stated. If the spacetime exterior to the horizon is empty, then Einstein's equations guarantee that $\sigma'=0$ so that $\sigma_H=\sigma_\infty=1$. More generally, the ratio $\sigma_H/\sigma_\infty$ depends on the matter content between the horizon and infinity. This is illustrated with some examples in Section V. It is for this reason that we refer to the standard definition as non-local.  It should be noted that Hayward and collaborators \cite{hayward} have used this non-locality to argue against the standard definition and motivate a local definition to be described below.


\subsection{A Non-Covariant Definition}
The first definition we choose to consider is due to Visser \cite{visser} and is defined in terms of the PG line element in (\ref{eq:PG}) in a slightly different format. Visser  writes the PG metric in the form
\be
ds^2 = -c(t,r)^2 dt^2 +\left( dr-v(t,r)dt \right)^2 + r^2 d\Omega^2 \label{eq:VisserPG}.
\ee
In our case, this corresponds to making the substitutions \(c(t,r)=\sigma(t,r)\) and \(v(t,r)=-\sigma(t,r) \sqrt{2GM(t,r)/r}\). The outer apparent horizon is then given by \(c=|v|\). Defining a quantity \(g(t)\) by
\be
g(t) \hat{\equiv} \frac{1}{2} (c^2-v^2)^\prime   \, ,
\ee
he then argues that the dynamical surface gravity is given by
\be
\kappa_V \hat{\equiv} \frac{g}{c} \hat{=} \frac{\sigma}{4GM} (1-2GM^\prime). \label{eq:visser}
\ee
By construction this coincides in form with the Killing definition \(\kappa_{Killing}\). Note that here and in what follows $\hat{=}$ implies that all quantities to the right are evaluated on the outer apparent horizon. The term ``outer horizon'' refers to the apparent horizon with the greatest areal radius at each spatial slice.

\subsection{Kodama Vector Definition}
Hayward {\it et al.} \cite{hayward} proposed a definition for spherically symmetric spacetimes based on the Kodama vector \cite{kodama}. They argued that this is a natural generalization because the Kodama vector field is parallel to the timelike Killing vector field if one exists and under certain circumstances coincides with it. The Kodama vector also mimics a  feature of the Killing vector that is key to the standard definition: it becomes null on the outer apparent horizon. The main disadvantage of this definition is that it is difficult to generalize the Kodama vector to non-spherically symmetric spacetimes (although see for example \cite{tung}).

The Kodama vector is given by
\be
K^{\mu} = \frac{1}{\sqrt{-\gamma}}\epsilon^{\mu \nu} \partial_{\nu}r  \,\, , \mu=0,1
\ee
where \(\epsilon^{\mu \nu}\) is the volume element associated with the horizon normal directions \cite{racz} and \(\gamma\) is the determinant of the 2D horizon normal subspace. In PG coordinates the Kodama vector is 
\be
K^{\mu} = \left(\frac{1}{\sigma},0,0,0\right).
\ee
Hayward {\it et al.} \cite{hayward} define the dynamical surface gravity to be:
\be
\kappa_H \equiv \frac{1}{\sqrt{-\gamma}}\epsilon_\mu{}^\alpha \partial_\alpha K^\mu=\frac{1}{2}g^{\mu \nu} \nabla_{\mu} \nabla_{\nu} r.
\ee
When evaluated on the outer apparent horizon in PG coordinates this yields
\be
\kappa_H \hat{=} \frac{1}{4GM} (1-2GM^\prime) + \frac{\dot{M}}{4M\sigma}, \label{eq:hayward}
\ee
where the overdot denotes differentiation with respect to the time coordinate. This definition is clearly local since it depends only on the mass function and its derivatives at the horizon. Although the last term depends on the lapse function it does so in the invariant combination $\sigma dt$. 

\subsection{Definitions Using Null Normals}
The next three definitions involve the use of the future/past pointing, ingoing and outgoing null normals. In PG coordinates, these vectors are calculated to be
\begin{gather}
l^{\mu} = \alpha \left( 1,\,-\sqrt{\frac{2GM}{r}}\sigma + \sigma,\,0,\,0 \right), \label{eq:lplus} \\
n^{\mu} = \beta \left( 1,\,-\sqrt{\frac{2GM}{r}}\sigma - \sigma,\,0,\,0 \right)
\end{gather}
where \(l^{\mu}\) is the (future) outgoing null vector and \(n^{\mu}\) is the (future) ingoing null vector for $\alpha > 0$ and $\beta > 0$. The scaling functions \(\alpha\) and \(\beta\) are functions of \(t\) and \(r\) that need to be specified in order to get a unique covariant definition of dynamical surface gravity. The cross normalization of the null vectors is:
\be
l^\mu n_\mu = -2\sigma^2 \alpha\beta.
\ee
We can now also calculate the outward and inward going null expansions in PG coordinates. They are, respectively:
\bea
\theta_l\equiv q^{\mu \nu} \nabla_{\mu} l_{\nu} = \frac{2\alpha\sigma\left(1-\sqrt{\frac{2GM}{r}}\right)}{r}, \label{eq:outexpansion} \\
\theta_n\equiv q^{\mu \nu} \nabla_{\mu} n_{\nu} = -\frac{2\beta\sigma\left(1+\sqrt{\frac{2GM}{r}}\right)}{r}.
\eea
where $q^{\mu \nu}$ is the projector onto the two-surface normal to $n^{\mu}$ and $l^{\mu}$. As expected for appropriate choices of the signs of $\sigma$, $\alpha$ and $\beta$ the outward expansion vanishes at $r=2GM$ and is positive if $GM/r \to0$, whereas the inward expansion is always negative. We now have the machinery at our disposal to examine various covariant definitions of surface gravity.

The first was proposed by Collins \cite{collins} in 1992. He derived a generalized first law that related the change of area of an apparent horizon to the change in what he interpreted as quasi-local energy. On the basis of this Collins identified the coefficient of the variation of the area as a generalized surface gravity, with the result:
\be
\kappa_{{Col}}\equiv \frac{n^\mu\nabla_\mu\theta_l}{\theta_n}\hat{=} \frac{\alpha\sigma}{4GM}\left(1- 2GM'\right)+ \frac{\alpha\dot{M}}{4M}.
\label{eq:collins}
\ee
The expression on the far right is valid in PG coordinates and is evaluated on an outer horizon $r=2GM$. Note that the normalization $\beta$ of the inward going normal has cancelled while the normalization of $l^a$ remains unspecified. This highlights a main issue in finding a useful definition of dynamical surface gravity using global vectors: the choice of parameterization for the respective null vectors, which translates to choosing the functions $\alpha$ and $\beta$. Collins did not, to the best of our knowledge, address this issue and we will argue below that one natural choice is $\alpha=1$. 

Fodor {\it et al.} \cite{fodor, nielsen1} proposed a definition using the outgoing null vector $l^\mu$ as follows:
\be
\kappa_F l^{\mu}= l^{\nu} \nabla_{\nu} l^{\mu} \label{eq:null def},
\ee
or alternatively
\be
\kappa_F = \frac{n_{\mu} l^{\nu} \nabla_{\nu} l^{\mu}}{l^\sigma n_\sigma}.
\label{eq:fodor def}
\ee
In order to specify $\alpha$ and $\beta$, they defined the cross-normalization to be \(l^{\mu}n_{\mu}=-1\) and fixed the ingoing null vector by requiring that it be everywhere affinely parameterized and normalized in an asymptotically static region such that $t^{\mu}n_{\mu} = -1$, where \(t^{\mu}\) is the time translational Killing vector normalized to unity at spatial infinity. These conditions result in a dynamical surface gravity that takes a particularly simple form in Eddington-Finkelstein coordinates, but is difficult to evaluate in the fully dynamical setting in PG coordinates. We defer a detailed analysis of this definition to future work. We instead retain (\ref{eq:fodor def}) with  the same cross normalization, but choose \( \alpha = 1 \) in (\ref{eq:lplus}). The resulting expression for the surface gravity is:
\be
\kappa_{PG} \hat{=} \frac{\sigma}{4GM} (1-2GM^\prime)+\frac{\dot{\sigma}}{\sigma}. \label{eq:kappa PG}
\ee
One feature of this choice is that \(l^{\mu} \rightarrow t^{\mu} \) as \(r \rightarrow r_+\) (where \(r_+\) is the outermost apparent horizon). Thus the outgoing null normal coincides with $\partial/\partial t$ on the outer apparent horizon, where $t$ is the PG time, which in turn coincides with the Killing vector in the static case. 

An alternative cross-normalization is provided by  Nielsen and Visser \cite{nielsen2} who choose \(l^{\mu}n_{\mu}=-2\) and the scaling function as \( \alpha = \beta = (\sigma)^{-1} \). This is their so called ``symmetric" choice of normalization and does not require an asymptotically flat spacetime. The expression for the surface gravity in PG coordinates is then given by
\be
\kappa_{NV} \hat{=} \frac{1}{4GM} (1-2GM^\prime). \label{eq:nielsen-visser}
\ee

Another definition based on null normals was recently given by Abreu and Visser \cite{abreu}. The starting point is the same as that of Fodor and Nielsen-Visser (i.e. Eq. (\ref{eq:null def})) but the normalization scheme is different. The outgoing null vectors are fixed in terms of a preferred time coordinate, namely the Kodama time. The corresponding expression is most simply formulated in Schwarzschild-like coordinates so this definition, like that of Fodor {\it et al.}, is not well adapted to PG coordinates. What is relevant for our purposes is the interesting averaging procedure used by Abreu and Visser. They first defined \( \kappa^\pm \) as the surface gravity evaluated via (\ref{eq:null def}), with the same cross normalization as Fodor {\it et al.}, associated with the outgoing future pointing and past pointing null normals \(l^{\mu}_\pm\), respectively. They then took the average of these to obtain:
\be
\kappa_{AV}=\frac{1}{2}(\kappa^+ +\kappa^-).
\ee
If we start instead from the $\kappa_{PG}$ in Eq. (\ref{eq:kappa PG}) and repeat this averaging procedure, we get a new definition of surface gravity (in PG coordinates):
\be
\kappa_{null} \hat{=} \frac{\sigma}{4GM}\left(1- 2GM'\right)+ \frac{\dot{M}}{4M}. \label{eq:PG null}
\ee
Remarkably Eq. (\ref{eq:PG null}) is precisely the same as Collins definition (\ref{eq:collins}) with the choice $\alpha=1$. We will henceforth refer to these two collectively as $\kappa_{null}$ for simplicity. Note that the expression in (\ref{eq:PG null}) differs from that of Hayward's only by a factor of $\sigma$: \(\kappa_{null} = \sigma \kappa_{H}\).  This is a crucial difference since it ensures that $\kappa_{null}$ reduces to the Killing definition in the stationary limit. 


\section{``Dirty" Black Holes}
By Birkhoff's theorem, any spherically symmetric solution to the vacuum Einstein equations can be put into the form (\ref{eq:PG}) with $M'=0$ and $\sigma'=0$.  However, in recent years, quantum gravity inspired modifications to Einstein's equations have lead to the consideration of static, spherically symmetric black hole spacetimes for which there is a non-zero effective stress-energy distribution everywhere, including the black hole exterior. The resulting lapse function \(\sigma=\sigma(r)\) depends non-trivially on the radial coordinate. These solutions are what are referred to as ``dirty" black holes \cite{dirty, visser2}. It is straightforward to calculate the Killing surface gravity for such solutions by first transforming to PG coordinates and using Eq. (\ref{eq:comparison}). We will do this in the following for a couple of representative examples.

\subsection{Stringy Charged Black Hole}
The stringy charged black hole \cite{stringy,stringy2} is a solution to a four-dimensional effective low energy theory obtained from string theory. The action contains an electromagnetic field and a metric both non-minimally coupled to a scalar dilaton field. The solution we wish to consider takes the form: 
\be
ds^2=-\left(1-\frac{2m}{R}\right)dt^2+\left(1-\frac{2m}{R}\right)^{-1}dR^2+R\left(R-\frac{Q^2e^{2\phi_0}}{m}\right)d\Omega^2.
\label{eq:stringy solution}
\ee
where $Q^2$ is the magnetic charge and $\phi_0$ is the asymptotic value of the dilaton. Note that the radial coordinate $R$ in (\ref{eq:stringy solution}) is not the areal radius. To put the metric in PG form, we first transform to the areal radius:
\be
r^2=R\left(R-\frac{Q^2e^{2\phi_0}}{m}\right).
\ee
This yields
\be
ds^2=-\left(1-\frac{2m}{R(r)}\right)dt^2+\left(1-\frac{2m}{R(r)}\right)^{-1}\left(\frac{dR}{dr}\right)^{2}  dr^2+r^2d\Omega^2,
\label{eq:stringy solution 2}
\ee
where \(R(r)\) is:
\be
R(r)=\frac{Q^2e^{2\phi_0}}{2m}+\left[r^2+\frac{Q^4e^{4\phi_0}}{4m^2}\right]^{1/2},
\ee
and 
\be
\left(\frac{dR}{dr}\right)=\frac{\sqrt{R(r)\left(R(r)-\frac{Q^4e^{4\phi_0}}{m}\right)}}{R(r)-\frac{Q^4e^{4\phi_0}}{2m}}.
\ee
Eq.~(\ref{eq:stringy solution 2}) is in the form (\ref{eq:schwarz static}) with 
\be
f(r)=\left(1-\frac{2m}{R(r)}\right) \left(\frac{dR}{dr}\right)^{-2} 
\ee
and
\be
\sigma(r)=\left(\frac{dR}{dr}\right).
\label{eq:sigma stringy}
\ee
There is one physical horizon located at $R_H=2m$, which corresponds to:
\be
r_H=\sqrt{2m\left(2m-\frac{Q^2e^{2\phi_0}}{m}\right)}.
\label{eq:stringy horizons}
\ee
Note that the horizon area goes to zero in the limit that $Q^2  e^{2\phi_0}=2m^2$, so that the horizon is singular in this limit \cite{stringy2}. 

Using the above formulae in the Killing definition (in fact all global definitions) yields: 
\be
{{\kappa_{Killing}\hat{=}\frac{1}{2}\sigma f'(r)= \frac{1}{4m}}}.
\ee
On the other hand, the local Hayward and Nielsen-Visser definitions give:
\be
\kappa_H=\kappa_{NV}\hat{=}\frac{1}{2}f'(r)=\frac{1}{4m}\frac{1-\frac{Q^2e^{2\phi_0}}{4m^2}}{\sqrt{1-\frac{Q^2e^{2\phi_0}}{2m^2}}}.
\ee
As anticipated the two classes of definition disagree when $Q^2\neq 0$. Moreover, the local definitions give a value for surface gravity that diverges in the singular limit $Q^2  e^{2\phi_0}=2m^2$, while the Killing definition yields a finite value.

\subsection{Quantum Corrected Black Hole}
This is a non-singular single horizon black hole spacetime derived in \cite{peltola} using a loop quantum gravity inspired quantization scheme similar to those applied to black hole interiors by a variety of authors \cite{ashtekar05, modesto06, boehmer07, pullin08}. The analytically continued spacetime describes a Einstein-Rosen type wormhole whose radius contracts to a minimum value determined by the quantum scale before re-expanding inside the black hole interior into an  asymptotically Kantowski-Sachs type spacetime. This scenario therefore realizes earlier proposals for ``universe creation'' inside black holes \cite{frolov90}.

The metric, for the quantum corrected black hole in Schwarzschild-like coordinates is
\be
ds^2=-\left( \epsilon \left(1-\frac{k^2}{r^2}  \right)^{1/2}  -\frac{2m}{r}  \right)dt^2+\left( \epsilon \left(1-\frac{k^2}{r^2}  \right)^{1/2}  -\frac{2m}{r}  \right)^{-1} \left( 1-\frac{k^2}{r^2}  \right)^{-1}dr^2,
\ee
where $m$ is the ADM mass of the black hole while $k$ is a new quantum scale. \(\epsilon=+1\) before the bounce and \(\epsilon=-1\) after the bounce. It was shown in \cite{peltola} by a straightforward change of coordinates that this analytic continuation produces a metric that is completely regular at the bounce. 

In order to calculate the surface gravity we note that for this metric:
\be
f(r)=\left( \epsilon \left(1-\frac{k^2}{r^2}  \right)^{1/2}  -\frac{2m}{r}  \right) \left( 1-\frac{k^2}{r^2}  \right)
\ee
while
\be
\sigma(r)= \left( 1-\frac{k^2}{r^2}  \right)^{-1/2}  \, .
\ee
The horizon radius, 
\be
r_H=(4m^2+k^2)^{1/2} \, \, ,
\ee
is always greater than the bounce radius $k$.
One can verify that
\be
\frac{df}{dr}\hat{=}\frac{2m}{4m^2+k^2},
\ee
where \(\epsilon=+1\) on \(r_H\) (before the bounce) and 
\be
\sigma \hat{=}\left(\frac{4m^2+k^2}{4m^2}   \right)^{1/2}.
\ee
 The Killing definition yields
\be
\kappa_{Killing}\hat{=}\frac{1}{2}\frac{1}{(4m^2+k^2)^{1/2}}.
\ee
Hayward's and Nielsen-Visser's definition, on the other hand
\be
\kappa_H=\kappa_{NV}\hat{=} \frac{m}{(4m^2+k^2)}.
\ee
 In this case  both expressions have roughly the same qualitative behaviour, but the  local form approaches the Schwarzschild value more rapidly as $k/m\to0$.



\section{Extremality Condition in the Dynamical Setting}
One condition that is often imposed is that the surface gravity of extremal black holes  be zero \cite{nielsen2, hayward}. A coordinate invariant definition of extremality that applies to dynamical horizons can be given as follows \footnote{Strictly speaking we are talking about generalized dynamical horizons \cite{ashtekar2}, since Ashtekar originally defined dynamical horizons to be purely spacelike \cite{ashtekar1}. }:
\begin{equation}
n^\mu\nabla_\mu \theta_l =0.
\label{eq:extremal condition}
\end{equation}
 Geometrically, a spherically symmetric dynamical horizon is extremal when the corresponding trapping surface is tangent to the inward going null vector. This in turn guarantees that the rate of change of the outward expansion along the inward null vector vanishes. Such extremal horizons can in principle form in dynamical settings \cite{booth08}.
 
In generalized PG coordinates a straightforward calculation yields:
\begin{equation}
n^\mu\nabla_\mu \theta_l = -\frac{\alpha\beta\sigma^2}{GMr} \left(1- 2 GM'+\frac{G\dot{M}}{\sigma}\right),
\end{equation}
so that dynamical extremal horizons satisfy: 
\be
1- 2 GM'+\frac{G\dot{M}}{\sigma}=0.
\ee
Clearly only the Hayward definition and $\kappa_{null}$ yield a surface gravity that generically vanishes for dynamical extremal horizons. 

One can also consider a slicing dependent notion of ``extremality'' as follows. For a spacelike slicing of an evolving black hole spacetime that is regular across future horizons, a given spatial slice can intersect the trapping horizon multiple times (See Figure (\ref{trapping})).  The instant of formation is defined as the time when the first spatial slice touches the trapping horizon. Future spatial slices necessarily contain an inner and an outer horizon. The former necessarily moves inward while the latter expands. At the instant of formation the inner and outer horizons coincide and in this sense the evolving black hole can be thought of as extremal.
The condition for this to occur is  that the trapping horizon be tangent to the spatial slice
In PG coordinates this can easily be worked out using Eq. (\ref{eq:outexpansion}) to yield the condition
\be
1-2GM'=0. \label{eq:PGextremal}
\ee
Note that in the case of multiple intersections, this condition is also satisfied on formation of any additional pair of horizons. 
Clearly, the definitions by Visser and Nielsen-Visser are the only ones that will generically vanish on formation in PG coordinates. This will be verified in the numerical simulations below.

\section{Numerical Method and Results}
We consider black hole formation via the collapse of a minimally coupled massless scalar field $\psi(r,t)$. The action is given by
\be
S=\frac{1}{16\pi G}\int d^4 x \sqrt{-g}  R(g) -\frac{1}{2}\int d^4 x |\nabla \psi|^2\, .
\ee
As derived in detail in \cite{ziprick},  the evolution equations for the scalar field $\psi$  and its canonical conjugate $\Pi_\psi$ in PG coordinates (\ref{eq:PG}) are:
\be
\dot{\psi} = \sigma \left( \sqrt{\frac{2GM}{r}} \psi^\prime + \frac{\Pi_\psi}{r^2}\right),
\label{psieom}
\ee

\be
\dot{\Pi}_\psi = \left[\sigma \left( r^2\psi^\prime + \sqrt{\frac{2GM}{r}} \Pi_\psi \right) \right]^\prime
\label{Pieom}
\ee
where we have absorbed a factor of $\sqrt{8\pi G}$ into $\psi$ in order to make it dimensionless.

The mass function $M$ is determined by the Hamiltonian constraint: 
\be
M^\prime =  \frac{1}{2}\left( \frac{\Pi_\psi^2}{r^2} + r^2(\psi^\prime)^2  \right)+\psi^\prime \Pi_\psi \sqrt{\frac{2G{ M}}{r}}.
\label{Meom}
\ee
With a slight abuse of terminology we henceforth refer to $M^\prime$ as the ``mass density.''
Consistency of the evolution equations in this gauge forces the lapse function to satisfy the constraint:
\be
\sigma^\prime + \frac{G \psi^\prime\Pi_\psi }{\sqrt{2G{M} r} } \sigma = 0.
\label{sigmaeom}
\ee
The above equations are evolved numerically with a slightly modified version of the code first used in \cite{ziprick}. 

The spatial grid resolution does not change with time, but does vary across the lattice. Near the origin, where high accuracy is needed, we use a fine mesh with a grid resolution of \(10^{-5}\). 
From just outside the origin to beyond the outermost apparent horizon the resolution is \(10^{-4}\). Finally, from outside the outermost horizon to the end of the lattice we use a grid resolution of \(10^{-2}\) which allows us to extend the lattice far enough to contain all the matter while reducing the CPU time for each simulation. This coarse resolution is still accurate enough to handle the dynamics and keep the code stable. A smoothing function is applied where the resolution changes. 

The time steps are determined using an adaptive refinement method. This method scans the spatial slice for each time step and refines the time spacing according to the condition
\be
\Delta t(t)=\text{MIN}_r \left \{ \frac{dt}{dr} \Delta r   \right \}, 
\ee
where \(dt/dr\) is the inverse of the local speed of either the ingoing or the outgoing null geodesic (whichever yields a smaller \(\Delta t\)). The time and spatial integrations are done using fourth order Runge-Kutta methods. For the run that we illustrate, there are  3000 time steps between the formation of the first set of apparent horizons (\(t=1.66\)) and the formation of the singularity (\(t=1.87\)). The code becomes unstable just before singularity formation, so the last 100 or so timesteps needed to be discarded.

\begin{figure}[ht!]
\begin{center}
\subfigure[Mass density]{
\includegraphics[width=0.47\linewidth]{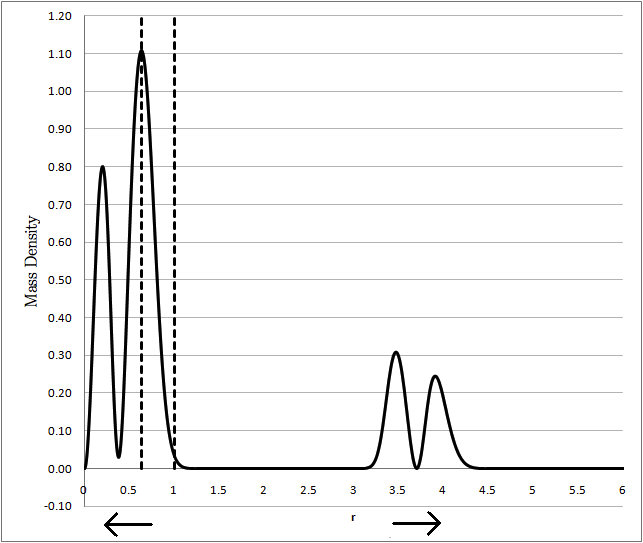}
\label{mass density}
}
\subfigure[Lapse function]{
\includegraphics[width=0.47\linewidth]{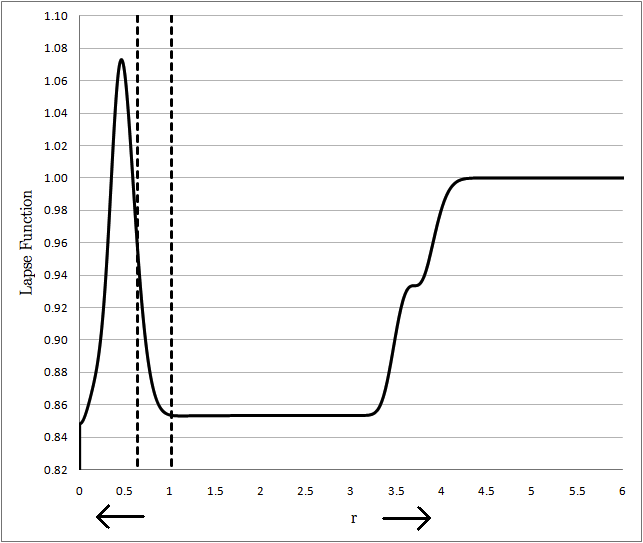}
\label{sigma}
}
\caption{Plots of mass density and lapse function as a function of radius after the formation of the trapping horizon at time \(t=1.73\). The dotted lines are the two apparent horizons (\(r_{\text{in}}=0.64\) and \(r_{\text{out}}=1.01\)). The lapse function has been normalized to unity at spatial infinity. The inner pulse is ingoing whereas the outer pulse is outgoing which is indicated with the arrows. They are sufficiently separated at this stage so that the region in between is approximately Schwarzschild. The mass function is constant in this region, while the lapse is spatially constant but increases with time. }
\label{collapse profiles}
\end{center}
\end{figure}

As initial data we choose a shell of matter described by a Gaussian scalar field:  
\be
\psi=A\exp\left[-\left( \frac{r-r_0}{B} \right)^2 \right].
\ee
Its conjugate momentum is initially set to zero. This choice of initial data actually corresponds to a linear combination of an ingoing and outgoing pulse. As seen in Fig. (\ref{collapse profiles}), by the time the initial horizon forms the two pulses are well separated and the mass function and the lapse function are spatially constant to a close approximation. The spacetime between the pulses is therefore very nearly Schwarzschild. By fine tuning the initial data one can in principle consider a pulse that is entirely ingoing, but the presence of an outgoing pulse at a large distance from the black hole turns out to be useful to the present discussion.

The parameters \(A\), \(B\) and \(r_0\) are the initial parameters and determine the size of the outer apparent horizon. For the particular results we are presenting here we used the initial parameters \( (A,\,B,\,r_0)=(0.185,\,0.3,\,3.0) \). This yields a black hole whose final mass (as determined by the mass function between the pulses) is about 0.51. The ADM mass of the spacetime, including the outgoing pulse, is about 0.65. 

\begin{figure}[ht!]
\begin{center}
\includegraphics[scale=0.5]{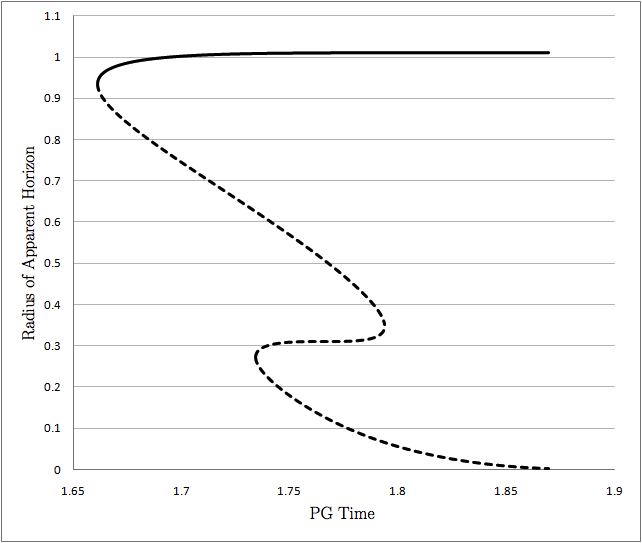}
\caption[]{Plot of location of outer apparent horizon (solid line) as a function of PG time. Note that between $t= 1.73$ and $t=1.79$ there are four apparent horizons. The singularity forms and the code terminates when the innermost horizon hits the origin at \(t=1.87\).}
\label{trapping}
\end{center}
\end{figure}

One of the main advantages of PG coordinates over null coordinates that are typically used for such simulations is that the slicings extend into the trapping region and the code can be run beyond horizon formation. Fig. (\ref{trapping}) maps out the trapping horizon as a function of PG time. One can see that the horizon forms initially at about t=1.66, and evolves into an inner and outer apparent horizon. As mentioned above, the former necessarily moves inward while the outer horizon grows until it reaches its final value after all the ingoing matter has fallen through. This occurs by about t=1.78. It is not unusual for additional pairs of apparent horizons to form, as happens in the black hole interior in this evolution at t=1.73. Once the innermost horizon reaches the origin, the singularity forms and the code terminates. It can be verified that the trapping horizon for our simulation is spacelike everywhere.

The lapse function has been normalized to unity at spatial infinity, i.e. outside the outermost pulse. As the system evolves, the lapse function in between the pulses can depend on both $r$ and $t$. After the pulses are well separated, Birkhoff's theorem requires the lapse between the pulses be spatially constant. However, the time dependence persists. Fig. (\ref{sigma before}) shows the lapse function on the outer apparent horizon as a function of PG time. It decreases rapidly while the horizon is growing. Once the matter has fallen through, it is spatially constant as expected but increases with time.  This is of course a coordinate dependent, kinematical effect that is a consequence of the presence of an outgoing outer pulse. The effect can easily be illustrated analytically by considering the outgoing pulse to be infinitely thin and using the Israel junction conditions (see for example \cite{poisson}). In this case, use of a single time coordinate on both sides of the shell while requiring continuity of the metric tangential to the null world line of the shell determines the discontinuity in the lapse across the shell be given by:
\be
\sigma_{in}(t)=\sigma_{out}(t)(1-2M/R_\Sigma(t))
\ee
where $R_\Sigma(t)$ is the radius of the shell at time $t$. If  $\sigma_{out}=1$, then  $\sigma_{in}$ asymptotes to one as the shell moves outwards. This is consistent with the observed increase in $\sigma$ on the outer horizon. In order to verify the details of the asymptotic behaviour one would have to run the code well past singularity formation. We have instead taken the simpler but equally effective route of doing a subcritical run in which all the matter reflects from origin and subsequently moves outward. Fig. ({\ref{fig:long_lapse}) illustrates the time dependence of the lapse at a radius $r_{obs}$ near the origin. Although the lapse exhibits the expected asymptotic approach to unity once the shell has escaped (i.e. $t > 3$) the convergence is surprisingly slow. 
 \begin{figure}[ht!]
\begin{center}
\includegraphics[width=0.8\linewidth]{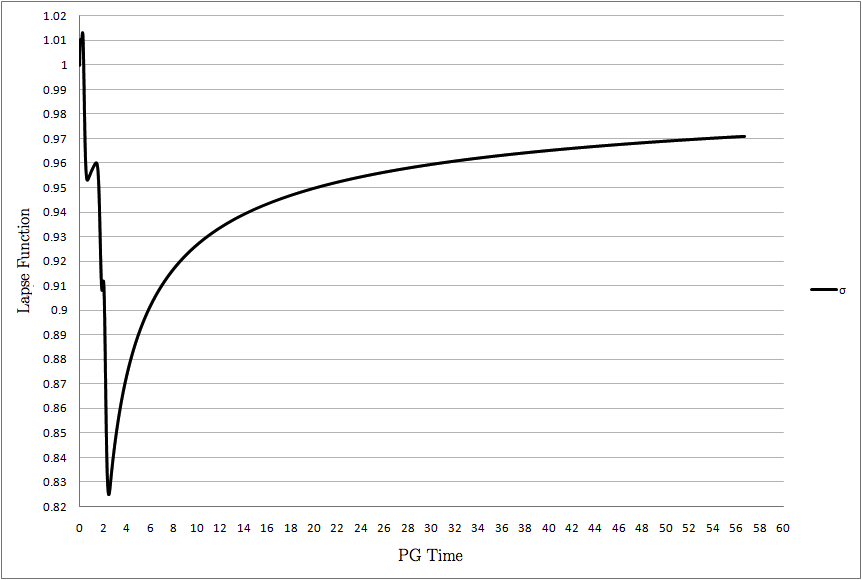}
\caption[]{Plot of the lapse function corresponding to subcritical collapse for initial parameters \((A,\,B,\,r_0)=(0.1,\,0.3,\,1.0)\) which yield an ADM mass of about 0.02. The lapse function is evaluated at the observation point of \(r_{obs}=0.77\). There are two pulses: an ingoing and an outgoing. The ingoing pulse falls below $r_{obs}$,  bounces and is reflected from the origin before escaping to spatial infinity just like the first outgoing pulse. The structure in the time dependence of $\sigma(t)$ between \(0\) and about \(2\) PG time is due to the passing of the inner pulse. After about 3 PG time units, all matter is outside $r_{obs}$ so that the spatial derivative of the lapse is approximately zero and the spacetime is very nearly flat.}
\label{fig:long_lapse}
\end{center}
\end{figure}

One can also choose a new time coordinate $\tilde{t}$ in which the lapse is constant in the region between the pulses. Fig. (\ref{sigma after}) plots the rescaled lapse $\tilde{\sigma}(\tilde{t})$  on the outer horizon in coordinates in terms of such a time coordinate. This corresponds to normalizing the local Killing vector $\partial/\partial \tilde{t}$ to one in the region just outside the horizon but well inside the outgoing pulse. The fact that $\tilde{\sigma}$ deviates slightly from unity at large PG times in Fig. (\ref{sigma after}) is a consequence of the fact that not quite all of the matter has yet fallen through the horizon at singularity formation, so that $\sigma'$ is not precisely zero.

\begin{figure}[ht!]
\begin{center}
\subfigure[Lapse Normalized at Infinity]{
\includegraphics[width=0.47\linewidth]{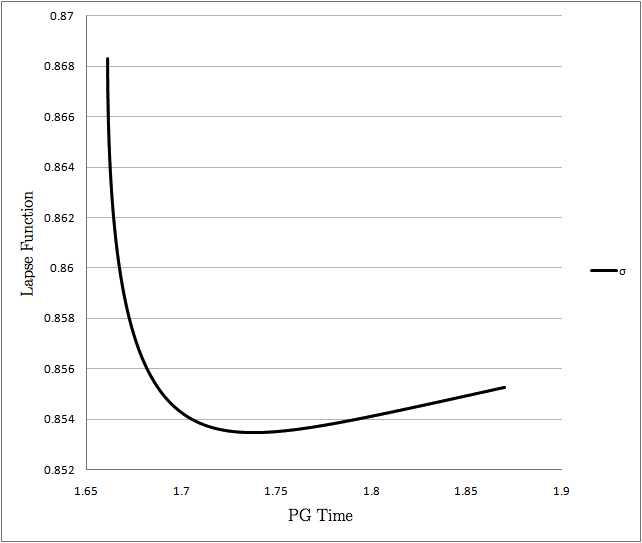}
\label{sigma before}
}
\subfigure[Lapse Normalized outside Horizon]{
\includegraphics[width=0.47\linewidth]{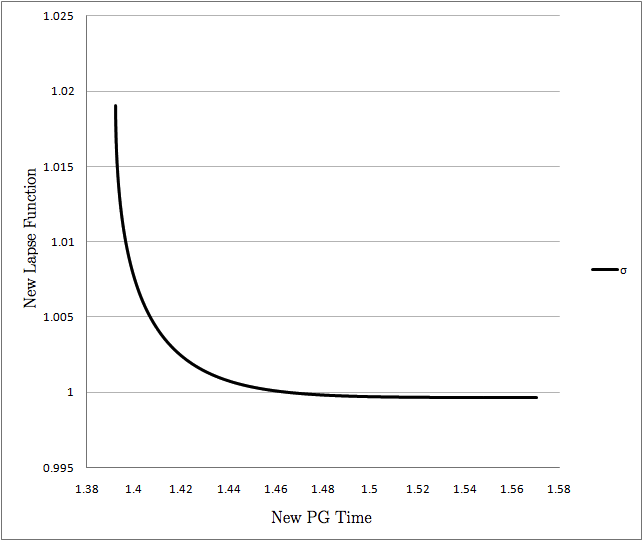}
\label{sigma after}
}
\caption[]{A plot on the left shows the lapse function at the horizon as a function of PG time normalized to be unity at spatial infinity.  The plot on the right shows the rescaled lapse function $\tilde{\sigma}$ (``New Lapse Function") at the horizon as a function of the rescaled time $\tilde{t}$ (``New PG Time").}
\label{both sigmas}
\end{center}
\end{figure}

The following graphs show the dynamical behaviour of the various definitions of surface gravity during the formation of the outer apparent horizon as it settles down to the final event horizon. 

\begin{figure}[ht!]
\begin{center}
\includegraphics[width=0.8\linewidth]{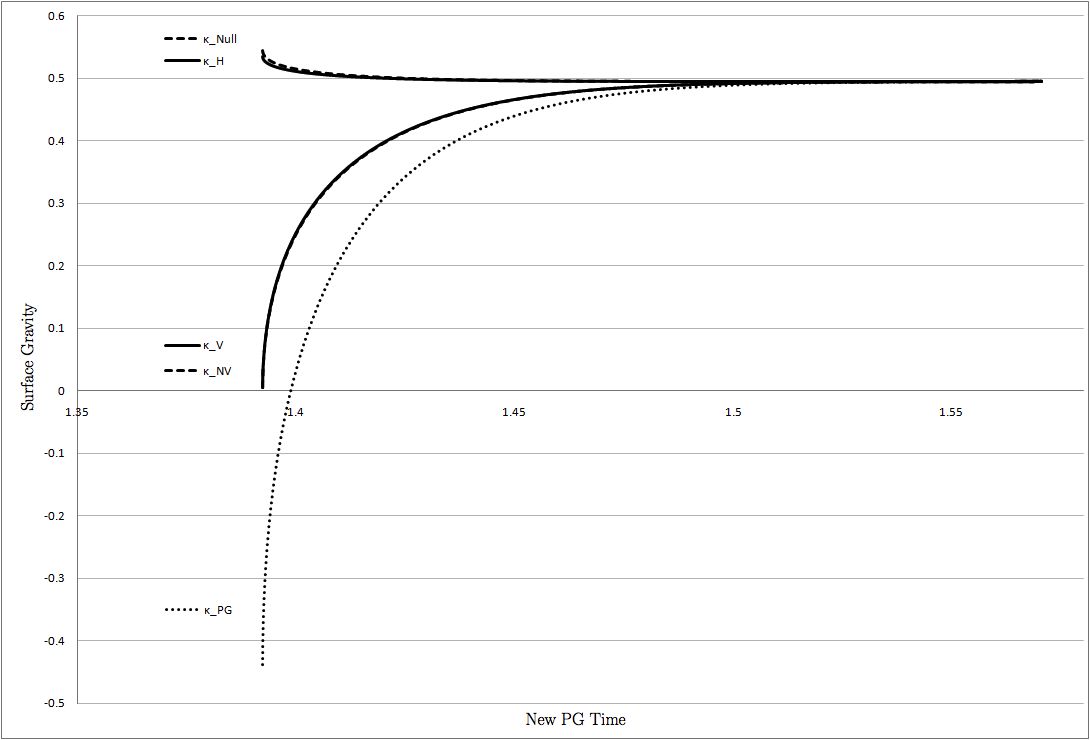}
\caption[]{Plot of various definitions of dynamical surface gravity on the outer horizon vs. new PG time with lapse chosen so that the timelike Killing vector is normalized to unity in between the two shells of matter. This is essentially equivalent to ignoring the outgoing pulse and treating the ingoing shell as isolated. In this case the asymptotic values of all definitions agree.}
\label{fig:sg_after}
\end{center}
\end{figure}

\begin{figure}[ht!]
\begin{center}
\includegraphics[width=0.8\linewidth]{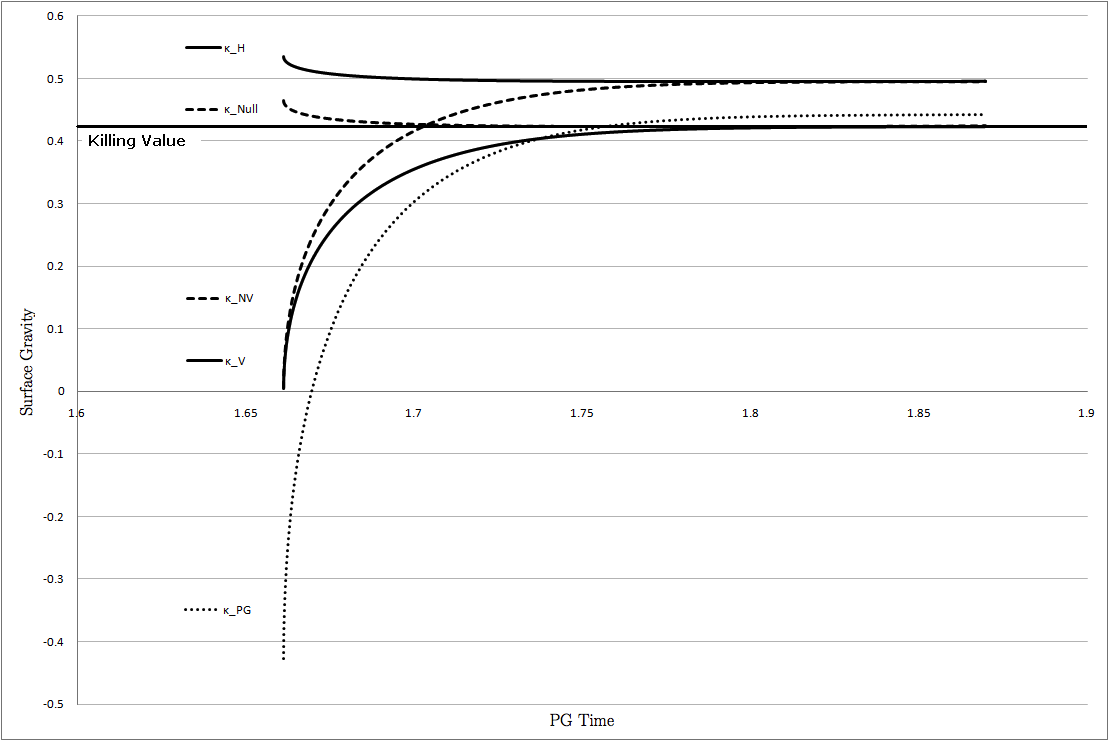} 
\caption[]{Plot of various definitions of dynamical surface gravity on the outer horizon vs. PG time with lapse chosen so that the timelike Killing vector is normalized to unity at spatial infinity. In this case the non-locality of some of the definitions is apparent: the asymptotic values do not agree even after the horizon has reached its final value.}
\label{fig:sg_before}
\end{center}
\end{figure}

Fig. (\ref{fig:sg_after}) calculates the surface gravity using the rescaled lapse $\tilde{\sigma}$, i.e. from the point of view of an observer in the approximately Schwarzschild region between the two pulses who is presumably unaware of the existence of the outgoing pulse.  From this perspective the spacetime settles down to that of an isolated Schwarzschild black hole. As expected all the definitions asymptotically converge to the same value of about 0.5, but the there is some variety in the rate of convergence. The definitions without time derivatives start at zero on formation and are practically indistinguishable because the lapse function  at the horizon does not change much during the evolution. $\kappa_{PG}$ on the other hand depends explicitly on the time rate of change of $\sigma$, which is relatively large and negative during the dynamical phase. The resulting surface gravity is therefore negative. The remaining two definitions start off slightly higher than the Killing value, because $\dot{M}$ is positive, and settle down very rapidly. 

Fig. (\ref{fig:sg_before}) on the other hand uses PG coordinates in which the lapse is normalized to unity at spatial infinity, i.e. outside the outgoing pulse, as demanded by the standard Killing definition. This does not change significantly the initial dynamical behaviour of the surface gravity, but it does affect the final equilibrium value. The Hayward and Nielsen-Visser definitions settle down to a different constant value because they are missing the overall factor of the lapse function which is not unity at the outer horizon even after all the matter has fallen through. As argued above one expects the lapse to asymptote at long times to one, so that the definitions will in this case ultimately agree, in distinction to what happens for static, dirty black holes.


\section{Summary and Conclusion}
We have presented an analytic and numerical study of several definitions of dynamical surface gravity. Our results are summarized in Table (\ref{table 1}). Only two of the definitions yield zero for extremal horizons. Moreover, as indicated in the last two columns, the different classes of definition  exhibit qualitatively different behaviour in the dynamical region. The PG definition (which contains an explicit $\dot{\sigma}$ in the definition) is negative on formation, which may be sufficient grounds to rule. The definitions with $\dot{M}$ start slightly above the equilibrium or final value, whereas the two definitions that had neither $\dot{\sigma}$ nor $\dot{M}$ start at zero on formation and take longer to converge.

As expected, all definitions agree for static horizons with vacuum exteriors. More generally (i.e. non-vacuum exteriors), even in the static case one must choose between locality (independence of matter content outside the horizon) and agreement with the Killing definition for all black holes with a global timelike Killing vector. This was illustrated with a couple of specific examples of dirty black holes and more graphically in our numerical analysis. The non-local definitions behave differently depending on whether one normalizes the Killing vector in the approximately Schwarzschild region directly outside the horizon (i.e. between the shells) or at spatial infinity. One somewhat unexpected result is that when the Killing vector is normalized at infinity the non-local definitions produce values for the surface gravity that increase with time even after all the matter has fallen through the horizon. This effect was attributed to the presence of the outgoing shell of matter between the observer at infinity and the event horizon and further highlights the non-locality of the standard definition.

One important question concerns the magnitude of the differences. Although we considered only massless scalar field collapse, our simulations nonetheless provide hints about the relative magnitude of the differences between the various definitions in more realistic astrophysical situations.  The massless scalar field dynamics is scale invariant in the sense that the amplitude of the initial pulse is dimensionless, while the units for the width $B$ and location $r_0$ can be specified arbitrarily. For the values of the parameters represented by Figs. 1 through 5, the mass of the shell is about 20\% of the final mass of the black hole, independent of the chosen scale.  Thus if one chooses the units of the initial parameters to be, say, ten kilometers, then the final horizon radius is about $5$ km, corresponding to a solar mass black hole, while the mass of the outgoing shell is about 0.2 of a solar mass. The difference between the local and non-local values of the final surface gravity, as represented by the asymptotic lines in Fig. 5, is about 20\% when the outgoing shell is about 4 horizon radii from the origin. Although the different values are expected to converge at very long PG times, the convergence is quite slow, as illustrated in Fig. (\ref{fig:long_lapse}).

\begin{table}[H]
\begin{center}
\begin{tabular}{|c|c||c|c|c|c|c|}
  \hline
  \multicolumn{7}{|c|}{{\bf Summary }} \\
  \hline
 \hline
  Name & Form in PG coords  &Covariant& Local&Killing & Zero for & Value on  \\[-10pt]
       &  on \(r=r_+ \)    &  definition?  & & &extremal? & formation       \\
  \hline 
  Visser & \large{ \(\frac{\sigma (1-2GM^{\prime})}{4GM}\) }&No & No & Yes & No& \( 0\) \\[5pt]
  Hayward & \large{ \( \frac{(1-2GM^{\prime})}{4GM}+\frac{\dot{M}}{4M \sigma} \) } & Yes & Yes& No & Yes&\(>\)final  \\[5pt]
   PG & \large{  \(\frac{\sigma (1-2GM^{\prime})}{4GM}+\frac{\dot{\sigma}}{\sigma}\) } & Yes & No &Yes & No&\(<0\) \\[5pt]
  Nielsen-Visser & \large{ \(\frac{(1-2GM^{\prime})}{4GM}\) } & Yes & Yes &No & No& \( 0\)\\[5pt]
  Null & \large{ \(\frac{\sigma (1-2GM^{\prime}) }{4GM}+\frac{\dot{M}}{4M}\) }& Yes & No &Yes  & Yes& \(>\)final  \\
  \hline
\end{tabular}
\end{center}
   \caption{Comparison of main properties of various definitions of dynamical surface gravity considered. The column labelled ``value on formation'' indicates whether the surface gravity starts off as zero, negative, or greater than the final (or equilibrium) value which is necessarily positive. In all cases studied the surface gravity is a monotonic function of PG time during the formation of the horizon.}
   \label{table 1}
\end{table}

It is important to note that while several of the surface gravity definitions we have investigated are motivated by and explicitly evaluated on apparent horizons, there is in principle nothing to 
stop us from evaluating the same quantities on null dynamical causal horizons such as the dynamical event horizon. In this case we would not expect any qualitative difference in our results.

Our results are intended to provide guidance with respect to a useful choice of definition for the surface gravity of dynamical horizons. If one requires the surface gravity of extremal horizons to be zero, then there are only two viable candidates: the Hayward definition and the null definition. For the initial data we used, both these definitions exhibited qualitatively similar behaviour during the dynamical phase of the horizon evolution. One must also choose between locality of the definition, i.e. independence from the matter content outside the horizon, and universal agreement with the standard Killing definition for static black holes, including ``dirty'' black holes.
In order  to proceed one must know what physical information black hole surface gravity is meant to provide.  For example, the Killing definition of surface gravity determines the tension in a massless string held by an observer at infinity suspending a weight just above the horizon. It is not surprising that such a measurement is sensitive to the mass distribution between the horizon and the observer. 

In a more modern context, it is the connection with black hole thermodynamics that makes surface gravity important. Most derivations yield the temperature of the black hole as measured at infinity, which is non-local in the sense given above. On the other hand, Hayward {\it et al.} \cite{hayward} have argued that in static cases at least, there exists a local definition of surface gravity that provides a local measure of the temperature of the horizon. 

Much work remains to be done before this important issue can be completely resolved. For example, in order to fully interpret dynamical surface gravity in terms of Hawking temperature, one needs a formalism for dealing quantitatively with the thermodynamics of dynamical horizons, i.e. the non-equilibrium thermodynamics of black holes. This is an even more fundamental and difficult problem.\\[5pt]

\noindent{\bf Acknowledgements}
We are grateful to S. Hayward for useful comments. G.K. and M.P. thank  E. Poisson and T. Taves for stimulating discussions and helpful insight. Special thanks to J. Ziprick for providing us with the original code as well as invaluable technical help. They also thank the Natural Sciences and Engineering Research Council of Canada and Career Focus of Canada, Manitoba Division for financial support.  A.B.N. is very grateful for generous support from the Alexander von Humboldt Foundation and hospitality at the Max
Planck Institute for Gravitational Physics in Potsdam-Golm. He also thanks the University of Winnipeg for its hospitality during the early stages of this collaboration.


\end{document}